\documentclass[sigconf]{acmart}
\usepackage{algorithm} 
\usepackage{algorithmic}
\usepackage{multirow}
\usepackage{mathrsfs}
\newcommand{\nop}[1]{}
\usepackage{marvosym}
\usepackage{tikz}
\usetikzlibrary{calc}
\usetikzlibrary{arrows.meta}

\copyrightyear{2020}
\acmYear{2020}
\setcopyright{acmcopyright}\acmConference[SIGIR '20]{Proceedings of the 43rd
International ACM SIGIR Conference on Research and Development in Information
Retrieval}{July 25--30, 2020}{Virtual Event, China}
\acmBooktitle{Proceedings of the 43rd International ACM SIGIR Conference on
Research and Development in Information Retrieval (SIGIR '20), July 25--30, 2020,
Virtual Event, China}
\acmPrice{15.00}
\acmDOI{10.1145/3397271.3401098}
\acmISBN{978-1-4503-8016-4/20/07}
\begin{document}
\fancyhead{}

\title{Incorporating User Micro-behaviors and Item Knowledge\\ into Multi-task Learning for Session-based Recommendation}
\author{Wenjing Meng,  Deqing Yang}
\authornote{Corresponding author.}
\email{
wjmeng18@fudan.edu.cn,   yangdeqing@fudan.edu.cn
}
\affiliation{
  \institution{School of Data Science, Fudan University}
  \streetaddress{No. 220 Handan Rd.}
  \city{Shanghai 200433}
  \country{China}
}

\author{Yanghua Xiao}
\email{shawyh@fudan.edu.cn}
\affiliation{
  \institution{School of Computer Science, Fudan University}
  \streetaddress{No. 220 Handan Rd.}
  \city{Shanghai 200433}
  \country{China}
}

\begin{abstract}
Session-based recommendation (SR) has become an important and popular component of various e-commerce platforms, which aims to predict the next interacted item based on a given session. Most of existing SR models only focus on exploiting the consecutive items in a session interacted by a certain user, to capture the transition pattern among the items. Although some of them have been proven effective, the following two insights are often neglected. First, a user's \emph{micro-behaviors}, such as the manner in which the user locates an item, the activities that the user commits on an item (e.g., reading comments, adding to cart), offer fine-grained and deep understanding of the user's preference. Second, the item attributes, also known as \emph{item knowledge}, provide side information to model the transition pattern among interacted items and alleviate the data sparsity problem. These insights motivate us to propose a novel SR model \textbf{MKM-SR} in this paper, which incorporates user \underline{\textbf{M}}icro-behaviors and item \underline{\textbf{K}}nowledge into \underline{\textbf{M}}ulti-task learning for \underline{\textbf{S}}ession-based \underline{\textbf{R}}ecommendation. Specifically, a given session is modeled on micro-behavior level in MKM-SR, i.e., with a sequence of item-operation pairs rather than a sequence of items, to capture the transition pattern in the session sufficiently. Furthermore, we propose a multi-task learning paradigm to involve learning knowledge embeddings which plays a role as an auxiliary task to promote the major task of SR. It enables our model to obtain better session representations, resulting in more precise SR recommendation results. The extensive evaluations on two benchmark datasets demonstrate MKM-SR’s superiority over the state-of-the-art SR models, justifying the strategy of incorporating knowledge learning.

\end{abstract}
\keywords{session-based recommendation, micro-behavior, multi-task learning, knowledge}

\maketitle 

\section{Introduction}
Recommender systems have played a very important role in many web applications including web search, e-commerce, entertainment and so on. %Many traditional recommendation algorithms, such as collaborative filtering (CF for short) \cite{sarwar2001itemcf,CF}, capture a user's preference mainly based on historical user-item interactions, thus suffer from the sparsity of interaction data \cite{Rendle2010Factorizing}. 
On many web sites, a user often exhibits a specific short-term intention \cite{Short_term_shopping}. The natural characteristics of a session data reflect a user's behavior pattern and current preference precisely. Therefore, modeling a user's current intention based on his/her behaviors in a recent session often results in satisfactory recommendation results, which is the basic principle of session-based recommendation (SR for short) \cite{Wang2019SurveySession}. As a representative class of sequential recommendation, SR systems aim to predict an item that would be interacted by a target user in the next interaction, based on the recent behaviors committed by the user in a session. 

%Due to the highly practical value, increasing research interests in this problem have risen. 
In order to achieve precise recommendations, an SR model generally uses a certain algorithm to leverage the sequential information in a session. The existing algorithms include Markov-based models and deep neural network (DNN for short) based models. For example, the basic idea of FPMC~\cite{Rendle2010Factorizing} is to calculate transition probability between the items in a session based on Markov chain. In recent years, recurrent neural networks (RNNs for short) have been applied in different ways to learn a user's dynamic and time-shifted preference~\cite{GRUrec,Jing2017Neural,SRhier,Liu2018STAMP}, exhibiting better performance than traditional methods. Although these deep SR models have been proven effective, there still exist some issues as follows.

\begin{figure}[htbp]
    \centering
    \includegraphics[width=9cm]{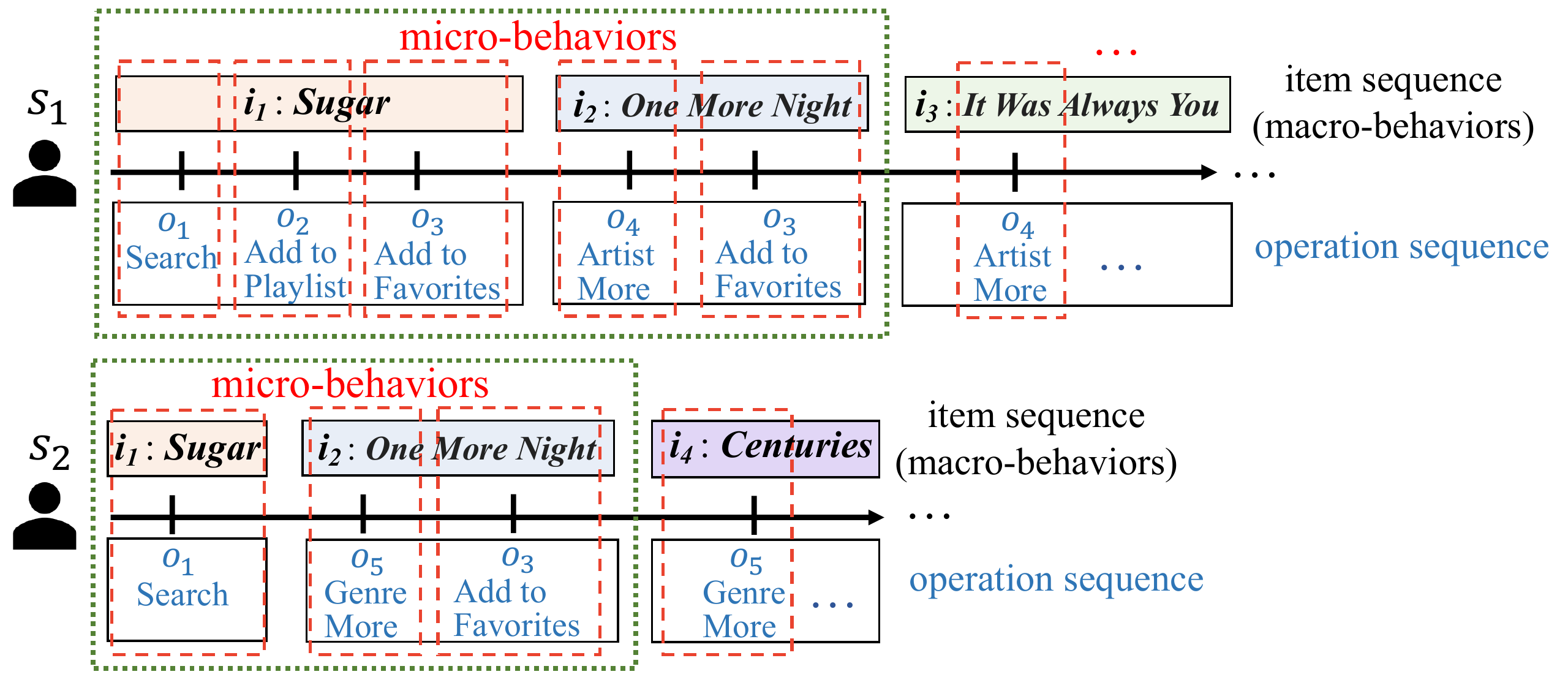}
    \vspace{-0.4cm}
    \caption{A toy example of user micro-behaviors (depicted by red dashed rectangles) in two sessions (depicted by green dashed rectangles) from a music site. Different operations may be committed on the same item sequence which reflect different intensions and preferences of a user on a fine-grained level. Therefore, modeling a session based on user micro-behaviors consisting of items and operations is helpful to achieve preciser SR.}
    \label{fig:micro_behaviors}
\vspace{-0.3cm}
\end{figure}

The first issue is the lack of exploiting user \emph{micro-behaviors}. Most of previous session-based models \cite{GRUrec,Jing2017Neural,Liu2018STAMP} only model a session from a macro perspective, i.e., regard a session as a sequence of items, without taking into account different operations of users. Even though a user interacts with the same item in a session, different operations committed on this item reflect the user's different intentions in this session and different preferences on this item. In this paper, we regard a user's specific operation on a certain item as a micro-behavior which has finer granularity and offers deeper understanding of the user than a macro-behavior of item-level, i.e., a user-item interaction. We use a toy example in Fig.\ref{fig:micro_behaviors} which was extracted from a real music dataset, to illustrate the significance of incorporating micro-behaviors into SR's session modeling.  

%In Fig.\ref{fig:micro_behaviors}, the user in session $s_1$ first searches song \textit{Sugar}, which is a funk-pop song of an American band \textit{Maroon 5}. Then, he adds it into his playlist and favorites in turn. Next, he clicks button \textit{Artist More} to get more \textit{Maroon 5}'s songs listed on a page, and then listens to song \textit{One More Night} on the page. After adding \textit{One More Night} into his favorites, he further listens to another song \textit{It Was Always You} of \textit{Maroon 5} through the \textit{Artist More} page. In session $s_2$, the user also first searches \textit{Sugar} and then clicks button \textit{Genre More} to get more funk-pop songs. Next, he also selects \textit{One More Night} from the genre page. After adding \textit{One More Night} into his favorites, he further listens to song \textit{'Centuries'} which also belongs to funk-pop but was sung by another artist. 
In Fig. \ref{fig:micro_behaviors}, the user in session $s_1$ first searches song $i_1$ which is a hip-pop song sung by a certain American band \textit{Maroon 5}. Then, he adds it into his playlist and favorites in turn. Next, he clicks button \textit{Artist More} to get more \textit{Maroon 5}'s songs listed on a page, and then listens to song $i_2$ on the page. After adding $i_2$ into his favorites, he further listens to another \textit{Maroon 5}'s song $i_3$ through the \textit{Artist More} page. Song $i_3$ belongs to a genre different from hip-pop although it is also sung by \textit{Maroon 5}. In session $s_2$, the user also first searches song $i_1$ and then clicks button \textit{Genre More} to get more hip-pop songs. Next, he also selects $i_2$ from the \textit{Genre More} page. After adding $i_2$ into his favorites, he further listens to song $i_4$ which also belongs to hip-pop but was sung by the singer other than \textit{Maroon 5}. 
Suppose that we already have the session information in the green dashed rectangles for the two sessions, we need to predict which item the user will interact with subsequently. According to traditional SR's paradigm of modeling a session only based on the historical item sequence, $s_1$ and $s_2$ are learned as the same representation since they both consist of song $i_1$ and $i_2$. Thus the same item may be recommended as the next interaction, which is not consistent with the observations in Fig. \ref{fig:micro_behaviors}. If we take user micro-behaviors which are depicted by red dashed rectangles rather than items to model the sessions, $s_1$ and $s_2$ have different representations. Based on such fine-grained session representations, $i_3$ will be recommended to the user in $s_1$ because the transition pattern between $i_2$ and $i_3$ (having the same singer) is the same as the one between $i_1$ and $i_2$.
%because $i_2\rightarrow i_3$ has the same micro-behavior transition (\textit{Artist More}) as $i_1\rightarrow i_2$. 
Similarly, $i_4$ will be recommended to the user in $s_2$.

\nop{Through modeling the fine-grained interaction (micro-behavior), we can better understand the users' intention and preference, which will result in improvement in predictive accuracy. 
 \begin{figure}[htbp]
    \centering
    \includegraphics[width=\linewidth]{behavior_embedding.png}
    \caption{The different representation of sessions}
    \label{fig:behavior embedding}
\end{figure}
}

The second issue is the insufficient utilization of item \emph{knowledge} towards the sparsity problem of user-item interactions. Since most of previous SR systems model a session only based on the interacted item sequence in the session, they can not learn session representations sufficiently when historical user-item interactions are sparse, especially for the cold-start items. Inspired by \cite{GANrec,KErec}, we can also import item attributes as side information which are generally distilled from open knowledge graphs (KGs for short) and recognized as item knowledge in this paper, to alleviate the data sparsity problem. %Although some recommendation models \cite{} have incorporated item knowledge to improve recommendation performance, they only use knowledge embeddings as pre-trained item embeddings. 
Furthermore, the item transitions in terms of micro-behavior can also be indicated by item knowledge. Recall the example in Fig. \ref{fig:micro_behaviors}, transition $i_2\xrightarrow{Artist\ More} i_3$ is indicated by two knowledge triplets <$i_1$, song-artist, \textit{Maroon 5}> and <$i_2$, song-artist, \textit{Maroon 5}>. Therefore, incorporating item knowledge is helpful for micro-behavior based modeling. In this paper, we regard such latent relationships among the items in a session as \emph{semantic correlations}.

%try to use item information, they only learning the knowledge representation in isolation, whether it's the item embedding initialization from a pre-trained knowledge graph or usage of the fixed embedding as a part of embedding concatenation. It's worth noting that, the co-occurrence of items in session and the relationships between items in knowledge graph have some consistency. For example, in $s_1$, the user listen to $i_1$ and $i_2$ successively through the operation \textit{'Artist More'}, and there are also corresponding relationships in the knowledge graph, <$i_1, song-artist, \textit{Maroon 5}>, <$i_2$, song-artist,\textit{Maroon 5}>. In this paper, we call the co-occurrence of items in session ‘the semantic information'. This phenomenon inspires us that we can learn the semantic information hidden in behavior sequences\cite{kang2019tdn} and the knowledge information unitedly to get more accurate item embedding.

To address above issues, we propose a novel SR model \textbf{MKM-SR} in this paper, which incorporates user \underline{\textbf{M}}icro-behaviors and item \underline{\textbf{K}}nowledge into \underline{\textbf{M}}ulti-task learning for \underline{\textbf{S}}ession-based \underline{\textbf{R}}ecommendation. In MKM-SR, a user's sequential pattern in a session is modeled on micro-behavior level rather than item-level. Specifically, a session is constituted by a sequence of user micro-behaviors. Each micro-behavior is actually a combination of an item and its corresponding operation. As a result, learning item embeddings and operation embeddings is the premise of learning micro-behavior embeddings, based on which a session's representation is generated. %We model the session in a micro-behavior level, as a micro-behavior can be regarded as the combination of a operation and item, the session sequence can be seen as a item sequence and a operation sequence. 
For this goal, we feed the operation sequence and the item sequence of a session into Gated Recurrent Unit (GRU for short) \cite{GRU} and Gated Graph Neural Network (GGNN for short) \cite{GGNN,wu2019session}, respectively. In this step, we adopt different learning mechanisms due to the different characteristics of operations and items. 
%thus we can get the micro-behavior embedding sequence, the concatenation of the output of the sequential item and operation embedding. Then we use attention layer to get the effective representation of the current session state. Next softmax the dot product of session representation and candidate items representation, we can get $\hat{y}$, the probability of each item to be recommended. Through backward the predict loss between the distribution of $\hat{y}$ and $y$, we can learn the session semantic information. 
What's more, we incorporate item knowledge to learn better item embeddings by TransH \cite{wang2014transh} which has been proven an effective KG embedding model on handling many-to-many/one relations. Unlike previous models using knowledge embeddings as pre-trained item embeddings \cite{DKN,KErec,GANrec}, we take knowledge embedding learning as an auxiliary task and add it into a multi-task learning (MTL for short) paradigm in which the major task is to predict the next interacted item. 
%We adopt TransH to learn the knowledge, and get the knowledge loss of knowledge triples. We add the predict loss and knowledge loss to get the multi-task loss, and these loss are weighted with task-dependant uncertainty. The multi-task learning enables the model to obtain better embeddings, as the auxiliary task provides additional evidence for the relevance and irrelevance of those features.
Our extensive experiments verify that our MTL paradigm is more effective than previous pre-training paradigm in terms of promoting SR performance.

In summary, our contributions in this paper are as follows:

    1. In order to improve SR performance, we incorporate user micro-behaviors into session modeling to capture the transition pattern between the successive items in a session on a fine-grained level, and further investigate the effects of different algorithms on modeling micro-behaviors.
    
    2. We incorporate item knowledge into our model through an MTL paradigm which takes knowledge embedding learning as an auxiliary (sub) task of SR. Furthermore, we validate an optimal training strategy for our MTL through extensive comparisons. 
    
    3. We provide deep insights on the rationales of our model's design mechanisms and extensive evaluation results over two realistic datasets (KKBOX and JDATA), to justify our model's superiority over the state-of-the-art recommendation models.

In the rest of this paper, we introduce related work in Section 2 followed by the detailed description of our model in Section 3. We display our extensive experiment results in Section 4 and conclude our work in Section 5.

\section{Related Work}
In this section, we provide a brief overview of the research related to our work in this paper.
%\vspace{-0.2cm}
\subsection{Session-based and Sequential Recommendation}
As a sub-task of sequential recommendation, the objective of SR is to predict the successive item(s) that an anonymous user likely to interact with, according to the implicit feedbacks in a session. In general, there are two major classes of approaches to leverage sequential information from users' historical records, i.e., Markov-based models and DNN-based models. In the first class, \cite{Shani2005AnMDP,Huang2009Markov} use Markov chains to capture sequential patterns between consecutive user-item interactions. \cite{ItemCF,Linden2003} try to characterize users' latest preferences with the last click, but neglect the previous clicks and discard the useful information in the long sequence. Rendel et.al proposed a hybrid model FPMC~\cite{Rendle2012Factorization}, which combines Matrix Factorization (MF for short) and Markov Chain (MC for short) to model sequential behaviors for next basket recommendation. A major problem of FPMC is that it still adopts the static representations for user intentions. Recently, inspired by the power of DNNs in modeling sequences in NLP, some DNN-based solutions have been developed and demonstrated state-of-the-art performance for SR. Particularly, the RNN-based models including LSTM and GRU, are widely used to capture users' general interests and current interests together through encoding historical interactions into a hidden state (vector). %These models can capture dynamic user preference over time and measure the likelihood of the next item.
As the pioneers to employ RNNs for SR, Hidasi et al. \cite{GRUrec} proposed a deep SR model which encodes items into one-hot embeddings and then feed them into GRUs to achieve recommendation. Afterwards, Jing et al. \cite{Jing2017Neural} further improved the RNN-based solution through adding extra mechanism to tackle the short memory problem inherent in RNNs. In addition, the model in \cite{Liu2018STAMP} utilizes an attention net to model user's general states and current states separately. This model takes into account the effects of users' current actions on their next moves explicitly. Besides, Hidasi et al. proposed \cite{hidasi2016parallel} and \cite{hidasi2018recurrent} to improve their model performance through adjust loss functions. More recently, the authors in \cite{wu2019session} adopted GGNN to capture the complex transition pattern among the items in a session rather than the transition pattern of single way. Although these models show promising performance for SR tasks, there is still room for improvement since they all neglect user micro-behaviors in sessions. \cite{liu2017multi,MBrec,rec18itemchain,hierachicalUserPro} are the rare models considering micro-behaviors. \cite{rec18itemchain} only models monotonic behavior chains where user behaviors are supposed to follow the same chain, ignoring the multiple types of behaviors. To tackle this problem, \cite{MBrec} and \cite{hierachicalUserPro} both adopt LSTM to model micro-behaviors. However, they ignored the different transition pattern between items and operations. In this paper, we adopt RNN and GNN simultaneously to model the micro-behaviors, which not only consider the differences of items and operations, but also keep the logic of operation order as mentioned in \cite{rec18itemchain,hierachicalUserPro}.

\vspace{-0.2cm}
\subsection{ Knowledge-based Recommendation}
Knowledge-based recommendation has already been recognized as an important family of recommender systems \cite{KGRec1}. Traditionally, the knowledge includes various item attributes which are used as the constraints of filtering out a user's favorite items. As more open linked data emerge, many researchers use abundant knowledge in KGs as side information to improve the performance of recommender systems. In \cite{EnRec} a heterogeneous information network (HIN for short) is constructed based movie knowledge, and then the relatedness between movies is measured through the volume of meta-paths. The authors in \cite{En2Rec} applied Node2Vec \cite{N2V} to learn user/item representations according to different relations between entities in KGs, but it is still collaborative filtering (CF for short) based method resulting in poor performance when user-item interactions are sparse. In recent years, many researchers utilized DNNs to learn knowledge embeddings which are fed into the downstream recommendation models. For example, Wang et al. proposed a deep model DKN \cite{DKN} for news recommendation, in which they also used a translation-based KG embedding model TransD \cite{TransD} to learn knowledge embeddings to enrich news representations. The authors in \cite{GANrec,KErec} utilized Metapath2Vec \cite{M2V} to learn knowledge embeddings which are used to generate the representations of users and items. Different with MKM-SR's multi-task learning solution, these models use knowledge embeddings as the pre-trained item embeddings. 
%The same authors also proposed CKE model \cite{CKE} which is fed with the knowledge distilled from texts and images besides structured knowledge (item attribute) in KGs. 
Another representative deep recommendation model incorporating knowledge is RippleNet \cite{RippleNet} where user representations are learned through an iterative propagation among a KG including the entities of items and their attributes. KGs have also been employed for sequential recommendation. For example, \cite{KMN} proposes a key-value memory network to incorporate movie attributes from KGs, in order to improve sequential movie recommendation. FDSA \cite{FDSA} is a feature-level sequential recommendation model with self-attentions, in which the item features can be regarded as the item knowledge used to enrich user representations. Unlike our model, these two KG-based sequential recommendation models model user behavior sequences on macro (item) level rather than micro level. What's more, our model incorporates item knowledge through an MTL paradigm which takes knowledge embedding learning as an auxiliary (sub) task of SR, guaranteeing the information is shared among user micro-behaviors and item attributes, thus representations are learned better.

\section{Methodology}
In this section, we introduce the details of MKM-SR including the related algorithms involved in the model. We first formalize the problem addressed in this paper, and then summarize the pipeline of MKM-SR followed by the detailed descriptions of each step (component). In the following introductions, we use a bold lowercase to represent a vector and a bold uppercase to represent a set, matrix or a cube (tensor).

\begin{figure*}[htbp]
    \centering
    \includegraphics[width=15.5cm]{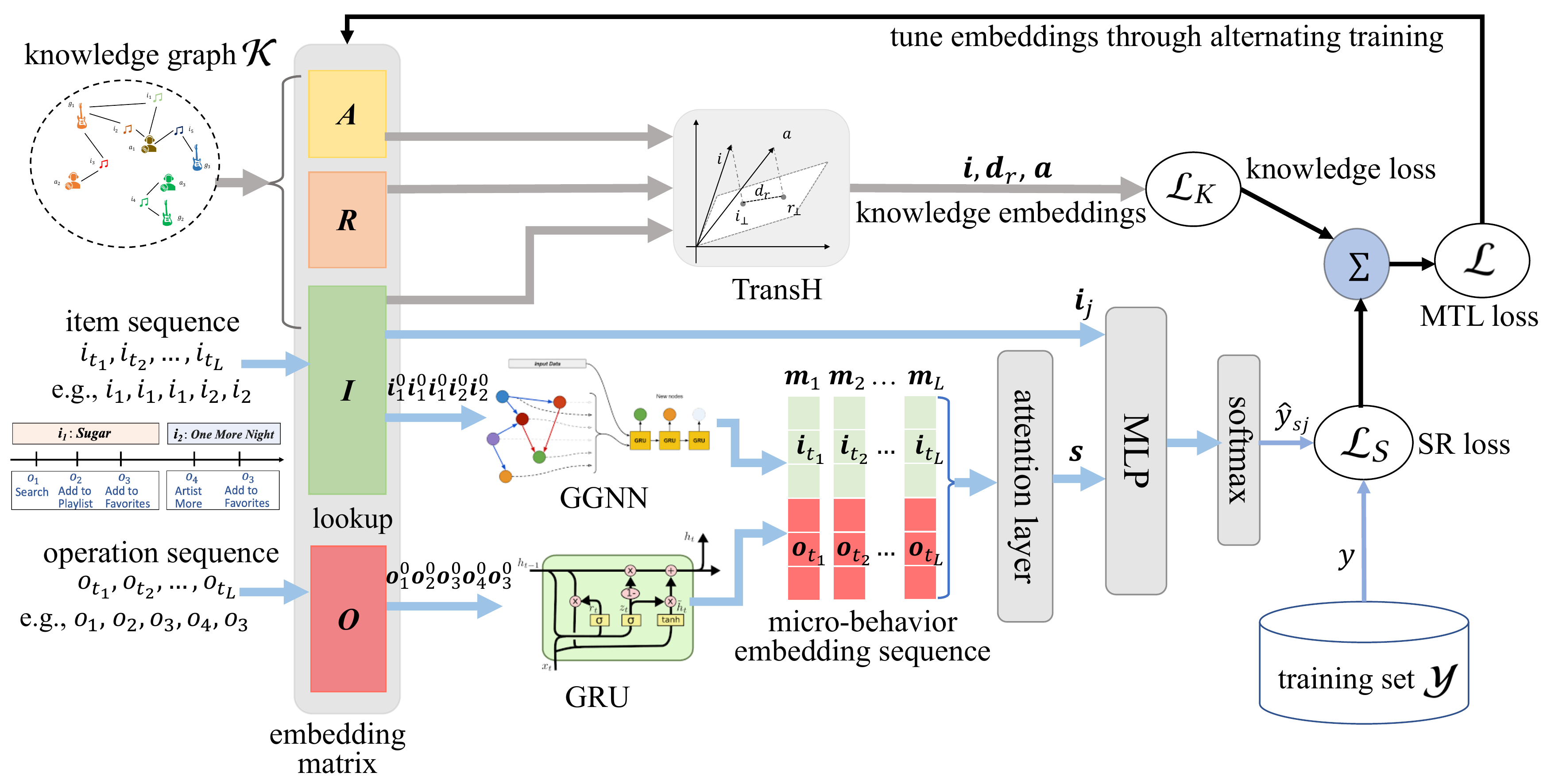}
    %\vspace{-0.2cm}
    \caption{The overall framework of our proposed MKM-SR. The arrows in the figure indicate data flows. At first, an item sequence and an operation sequence are extracted from a given session simultaneously, and then fed into GGNN and GRU to learn item embeddings and operation embeddings, respectively. These two types of embeddings assemble a sequence of micro-behavior embeddings which are used to generate the session's representation $\mathbf{s}$. The final score $\hat{y}_{sj}$ is computed by an MLP followed by softmax operation. Furthermore, the knowledge embeddings learned by TransH are incorporated into a multi-task learning loss function to learn better item embeddings resulting in superior SR.}
    \label{fig:framework}
\end{figure*}

\subsection{Problem Definition}
%In a session-based recommender, we use $S=\{s_1,s_2,\dots,s_{N_S}\}$ denote the set of sessions, and $ I = \{i_1 ,i_2,\dots,i_{N_I}\}$ denote the set all items. Since a anonymous session sequence $s_i$ can be represent by a list of micro-behaviors: $s_i= [m_1^{s_i},m_2^{s_i},\dots,m_L^{s_i}]$, and the $m_j^{s_i}$ is the micro-behavior at the $j-th$ entry of session $s_i$. The micro-behavior $m_i=(o_p,i_q)$ means that a user interact with  $i_q$ in the $o_p$ operation type. 

In this paper, since we focus on user micro-behaviors rather than interacted items in sessions in this paper, we first use $\{m_1, m_2, ..., m_L\}$ to denote the micro-behavior sequence in a session $s$ where $L$ is the length of the sequence. Specifically, $m_t(1\leq t\leq L)$ is the $t$-th micro-behavior which is actually a combination of an item and its corresponding operation. Furthermore, the item knowledge we incorporate into MKM-SR is represented by the form of triplet. Formally, a knowledge triplet $<i, r , a>$ represents that $a$ is the value of item $i$'s attribute $r$ which is often recognized as a relation. For example, <\textit{Sugar}, song-artist, \textit{Maroon 5}> describes that \textit{Maroon 5} is the singer of song \textit{Sugar}. MKM-SR is trained with the observed user micro-behaviors in sessions and obtained item knowledge.

The goal of our SR model is to predict the next interacted item based on a given session. To achieve this goal, our model is fed with the given session $s$ and a (next interacted) candidate item $j$ to generate a matching score (probability) $\hat{y}_{sj}$ between $s$ and $j$. With $\hat{y}_{sj}$, a top-$k$ ranking list can be generated for a given session (one sample). In general, the item with the highest score will be predicted as the next interacted item.
%Let $\mathbf{\hat{y}}=\{\hat{y_1}, \hat{y_2},\dots,\hat{y_{N_I}}\}$ denote the output score vector, where $\hat{y_j}$ corresponds to the score of item $i_j$. The item with top-K values in $\mathbf{\hat{y}}$ will be the candidate items for recommendation.

%In this section, we first give an overview of our work. Then we will describe the representation of micro-behavior. After that, we will discuss how to embed the session correlation and item attribute in detail and give some discussions on the proposed model. Finally, we will describe the method our model to unit the two related work organically and the training phase.
%\vspace{-0.2cm}
\subsection{Model Overview}
The framework of MKM-SR is illustrated in Fig. \ref{fig:framework}. In model training, besides the operations and items involved in training samples, item knowledge is also input into MKM-SR to learn better operation embeddings and item embeddings, then better session representations are also obtained which are crucial to compute precise scores.
%MKM-SR takes the session sequences and knowledge triples as input, and output the predicted probability that items to be interacted with next for a given session sequence. 

For more clear explanation, we reversely introduce the pipeline of MKM-SR from the right side of Fig. \ref{fig:framework}. For a given session $s$ and a candidate item $j$, the final score $\hat{y}_{sj}$ is computed by a multi-layer perceptron (MLP for short) fed with $s$'s representation and $j$'s embedding which are both a vector of $d$ dimensions. Specifically, $s$'s representation $\mathbf{s}$ is obtained by aggregating a group of micro-behavior embeddings. Given that different objects in a session (sequence) have different levels of priority on representing this session, we adopt a soft-attention mechanism \cite{softAtt} to generate $s$'s global representation which reflects a user's long-term preference. Furthermore, a micro-behavior embedding is the concatenation of of its item embedding (green rectangles in Fig. \ref{fig:framework}) and operation embedding (red rectangles in Fig. \ref{fig:framework}) since we believe an operation and an item have different roles to represent a user's micro-behavior. The item embedding and the operation embedding in a micro-behavior embedding are learned respectively by different models, i.e., GGNN and GRU. The reason of such manipulations will be explained subsequently. The GGNN and the GRU in MKM-SR are fed with an item sequence and an operation sequence respectively, both of which have the same length as the micro-behavior sequence, i.e., $L$.
%As each micro-behavior is a tuple of an operation and an item and they have different roles, we model the item sequence and operation sequence in GGNN and GRU respectively(later we will discuss the reason thoroughly), to get the sequential representation of the session. Then we concatenate the item and operation embedding to get the real-time micro-behavior state. Consider information in these embedding may have different levels of priority, we further adopt the soft-attention mechanism to get the global session representation. After obtained the embedding of each session, we compute the score $\hat{y_j}$ for each candidate item $i_j\in I$ by softmax the dot product of  the item embedding $\mathbf{i_j}$ and session representation $\mathbf{s}$.

%The training phase is a little more complex, where the data will flow through all of the framework. Firstly, after prediction, we can get the cross-entropy of the prediction and the ground truth, which we call 'session predict loss'. Besides, 
As we mentioned before, the item knowledge is helpful to discover the semantic correlations between the items in a session. As a result, we add knowledge learning as an auxiliary task into an MTL paradigm through designing a weighted sum of different loss functions. Specifically, we import TransH's \cite{wang2014transh} loss function as the loss of our knowledge learning, since it is a KG embedding model which can model many-to-many/one relations effectively. In addition, we adopt \emph{alternating training} \cite{ren2015faster} strategy for training MKM-SR.

\subsection{Encoding Session Information}\label{sec:session}
In this subsection, we introduce how to obtain the representation of a given session which is crucial for our model to compute the final score $\hat{y}$. Based on the basic principle of SR \cite{GRUrec,Jing2017Neural}, the premise of obtaining a session representation is to learn each object's embedding in the session. In the setting of this paper, an object in the sequence of a session is a micro-behavior. According to our definition of micro-behavior, a micro-behavior is the combination of an item and an operation committed on the item. In MKM-SR, we first learn item embeddings and operation embeddings separately, and then concatenate an item embedding and an operation embedding as the embedding of the micro-behavior, rather than directly learn a micro-behavior embedding as whole. Our experiment results shown in the subsequent section verify that our solution is better than the latter. We adopt such solution due to the following intuitions.

\subsubsection{Intuitions of Embedding Learning}
We argue that \emph{item sequences and operation sequences have different effects on session modeling, and exhibit different transition patterns}. For the item sequence of a session, its transition pattern is actually more complex than the single way transitions between successive items which were captured by previous RNN-based sequential models \cite{GRUrec,Jing2017Neural,SRhier}. In other words, not only the subsequent items are correlated to the preceding ones in a sequence, but also the preceding items are correlated to the subsequent ones. It is also the reason that a user often interacted with an item that he/she has already interacted with before. Obviously, such transition pattern relies on the bidirectional contexts (preceding items and subsequent items) rather than the unidirectional contexts, which can be modeled by a graph-based model rather than a sequential model of single way such as GRU. Consequently, inspired by \cite{wu2019session} we adopt GGNN to model item sequences to obtain item embeddings in MKM-SR.

Although the operations committed by a user in a session also assemble a sequence, their transition pattern is different with the one of item sequences. Therefore, GGNN is not appropriate to model operation sequences due to the following reasons. First, the unique types of operations are very limited in most platforms. One operation may recur in a sequence with big probability, resulting in that most nodes (operations) have similar neighbor groups if we convert operation sequences into a directed graph. Thus, most operation embeddings learned through applying GGNN over such a graph are very similar, which can not well characterize the diversity of a user's preference. On the other hand, the transitions between two successive operations often demonstrate a certain of sequential pattern. For example, a user often adds a product to cart after he/she reads its comments, or purchases the product after he/she adds it to cart. Therefore, we adopt GRU rather than GGNN to learn operation embeddings. Next, we introduce the details of learning item embeddings and operation embeddings in turn.

%In a short session, users' intention is relatively single and focused, that is to say, the items co-occur in a session have the similarity. And according to previous studies , the orders in which users pick up the items have local randomness, affected by the layout of the page, the items order and other factors. What's more, in reality, the user may interact with item previously interacted with again. Thus, when model the item sequential state, it's necessary to take all the transitions in the session into consideration. 

%Gated Graph Sequence Neural Network(GGSNN) is effective way to model the complex transitions, which can learn the representations of the directed-graph internal state. Comparing the other graph model only outputs a single graph-level output, this method can output the sequence. Through this, the features learn the graph and still keep the flexibility to be reprocessed.

\subsubsection{Learning Item Embeddings}
In order to learn item embeddings by GGNN, we should convert an item sequence into a directed graph at first. 

Formally, given a micro-level item sequence $S^i=\{i_{t_1}, i_{t_2}, ... ,i_{t_L}\}$ in a session in which each object is the item in a micro-behavior, the corresponding directed graph is $\mathcal{G} = (\mathcal{V},\mathcal{E})$. In $\mathcal{G}$, each node represents a unique item in $S^i$, and each directed edge $(i_{t_{k-1}},i_{t_k}) \in \mathcal{E} (2\leq k\leq L)$ links two successive items in $S^i$. Please note that an item often recurs in a session. For example, the item sequence of session $s_1$ in Fig. \ref{fig:micro_behaviors} is $\{i_1, i_1, i_1, i_2, i_2\}$. As a result, $|\mathcal{V}|\leq L$ and self-loops exist in $\mathcal{G}$ if $i_{t_{k-m}}=i_{t_k},(1\leq m \leq k-1)$. To better model $\mathcal{G}$, we further construct it as a weighted directed graph. The normalized weight of edge $(i_{t_{k-1}},i_{t_k})$ is calculated as the occurrence frequency of $\{i_{t_{k-1}},i_{t_k}\}$ in $S^i$ divided by the frequency that item $i_{t_{k-1}}$ occurs as a preceding item in $S^i$. 

In the initial step of GGNN, for a given item node $v$ in $\mathcal{G}$, its initial embedding $\mathbf{i}^0_v\in \mathbb{R}^d$ is obtained by the lookup of item embedding matrix, and used as its initial hidden state $\mathbf{h}^0_v$. Based on the basic principle of iterative propagation in GNN \cite{GNN}, we use the hidden state in the $h$-th ($1\leq h\leq H$) step as $v$'s item embedding after $h$ steps, i.e., $\mathbf{i}^h_v=\mathbf{h}^h_v$. Since $\mathcal{G}$ is a weighted directed graph, we use $\mathbf{A}^+\in\mathbb{R}^{|\mathcal{V}|\times|\mathcal{V}|}$ and $\mathbf{A}^-\in\mathbb{R}^{|\mathcal{V}|\times|\mathcal{V}|}$ to denote $\mathcal{G}$'s incoming adjacency matrix and outgoing adjacency matrix, respectively. The entries of these two adjacency matrices are edge weights indicating the extent to which the nodes in $\mathcal{G}$ communicate with each other. Then, $\mathbf{h}^h_v$'s is computed according to the following update functions,
\begin{equation}
     \begin{split}
      \mathbf{a}_v^h &= (\mathbf{A}^+_{v:}+\mathbf{A}^-_{v:}) [\mathbf{h}^{h-1}_1, \mathbf{h}^{h-1}_2, ..., \mathbf{h}^{h-1}_{|\mathcal{V}|}]^{\top} + \mathbf{b}\\
        \mathbf{z}_v^{h} &= \sigma(\mathbf{W}_{az}\mathbf{a}_v^{h} + \mathbf{W}_{hz}\mathbf{h}_v^{h-1})\\
        \mathbf{r}_v^{h}& = \sigma(\mathbf{W}_{ar}\mathbf{a}_v^{h} + \mathbf{W}_{hr}\mathbf{h}_v^{h-1})\\
        \mathbf{c}_v^{h} &= tanh\big(\mathbf{W}_{ac}\mathbf{a}_v^{h} + \mathbf{W}_{hc}(\mathbf{r}_v^{h}\odot\mathbf{h}_v^{h-1})\big) \\
        \mathbf{h}_v^{h} &= (1-\mathbf{z}_v^{h})\odot\mathbf{h}_v^{h-1} + \mathbf{z}_v^{h}\odot\mathbf{c}_v^{h}\\        
    \end{split}
    \label{eq:GGNN}
\end{equation}
where all bold lowercases are the vectors of $d$ dimensions and all $\mathbf{W}$s are $d\times d$ matrices. $\sigma$ is Sigmoid function and $\odot$ is element-wise multiplication. In addition, $\mathbf{r}_v^{h}$ and $\mathbf{z}_v^{h} $ are reset gate and update gate respectively. As described in Eq. \ref{eq:GGNN}, the hidden state in the $h$-th step for item $v$, i.e., $\mathbf{h}_v^{h}$, is calculated based on its previous state $\mathbf{h}_v^{h-1} $ and the candidate state $\mathbf{c}_v^{h} $. After $H$ steps, we can get the learned embeddings of all item nodes in $\mathcal{G}$, based on which the item embeddings in $S^i$ are obtained as
\begin{equation}
    \mathbf{\widetilde{I}} = [\mathbf{i}_{t_1}, \mathbf{i}_{t_2}..., \mathbf{i}_{t_L}]^\top=[\mathbf{h}^{H}_{t_1}, \mathbf{h}^{H}_{t_2}..., \mathbf{h}^{H}_{t_L}]^\top\in\mathbb{R}^{L\times d}
\end{equation}

According to $\mathcal{G}$'s construction, given a session, an item has only one learned embedding no matter whether it recurs in the sequence. Consequently, $\mathbf{\widetilde{I}}$ may have recurrent item embeddings. If an item occurs in multiple sessions, they may have different learned embeddings since different sessions correspond to different $\mathcal{G}$s.

\subsubsection{Learning Operation Embeddings}
%Actually, the operations works like adding offsets to the item. Different offsets can model the different degrees of preference on the item or the different reasons why the user interact with the item. Though we can also use the operation sequence to construct a operation directed graph, the graph neural networks don't work well in this scenario. The reason is that, for a platform, the types of operations is very limited, the co-occurrence of operations is so common that all operations are neighbors. As GGSNN will propagate the neighborhood information, the differences between operations will be smoothed, and it will not play a role in reflecting different user preferences for different operations. So, we just adopt GRU, the traditional Recurrent Neural Network, to learn the sequential operation state.

Due to the aforementioned reasons, we adopt a GRU \cite{GRU} fed with operation sequences to learn operation embeddings. GRU is an improved version of standard RNN to model dynamic temporal behaviors, which aims to solve the vanishing gradient problem. %It uses two vectors to decide what information should be passed to the output. The two vectors are called update gate $\mathbf{z}^{h}$and reset gate $\mathbf{r}^{h}$, and are trained to control the input $\mathbf{i}^{h}$ in the $h$-th step and the previous hidden state $\mathbf{h}^{h-1}$. 

Formally, we use $S^o=\{o_{t_1}, o_{t_2}, ... ,o_{t_L}\}$ to denote an operation sequence fed into our GRU. For an operation $o_{t_k} (1\leq k\leq L)$ in $S^o$, its initial embedding $\mathbf{o}_{t_k}^0$ is also obtained by the lookup in operation embedding matrix. Then, its learned embedding $\mathbf{o}_{t_k}$ is the hidden state (vector) in the $k$-th step output by GRU, which is calculated based on $\mathbf{o}_{t_k}^0$ and the hidden state in the ($k$-1)-th step as follows,
\begin{equation} \label{eq:gru}
       \mathbf{o}_{t_k}= \mathbf{h}_{t_k} = GRU(\mathbf{h}_{t_{k-1}},\mathbf{o}_{t_k}^0;\Phi_{GRU})
\end{equation}
where $GRU(\cdot)$ represents the calculation in one GRU unit, and $\Phi_{GRU}$ denotes all GRU parameters. In fact, $\mathbf{h}_{t_{k-1}}$ is $o_{t_{k-1}}$'s learned embedding. To calculate $\mathbf{h}_{t_1}$, we set $\mathbf{h}_{t_0}=\mathbf{o}^0_{t_1}$.
%The output $ \mathbf{h}_o^{(t)} $ can be regarded as the state of the $t-th$ operation and is adjusted according to the whole sequence before it.
Thus, we obtain the learned embeddings of all operations in $S^o$ as 
\begin{equation}
    \mathbf{\widetilde{O}} = [\mathbf{o}_{t_1}, \mathbf{o}_{t_2}, ..., \mathbf{o}_{t_L}]^\top\in\mathbb{R}^{L\times d}
\end{equation}

Please note that an operation may also recur as an item in the sequence. According to GRU's principle, an operation recurring in an operation sequence has multiple different embeddings. For example, the operation sequence of $s_1$ in Fig. \ref{fig:micro_behaviors} is $\{o_1, o_2, o_3, o_4, o_3\}$. $o_3$'s learned embedding in the third position is different to its learned embedding in the fifth position in the sequence. As a result, $\mathbf{\widetilde{O}}$ has no recurrent embeddings, which is different to $\mathbf{\widetilde{I}}$.

Then, we concatenate $\mathbf{\widetilde{O}}$ and $\mathbf{\widetilde{I}}$ to obtain the embeddings of the $L$ micro-behaviors in the given session as shown in Fig. \ref{fig:framework}. So we have
\begin{equation}
     \mathbf{\widetilde{M}} = [\mathbf{m}_1, \mathbf{m}_2, ..., \mathbf{m}_L]^\top=[\mathbf{i}_{t_1}\oplus\mathbf{o}_{t_1}, \mathbf{i}_{t_2}\oplus\mathbf{o}_{t_2}, ..., \mathbf{i}_{t_L}\oplus\mathbf{o}_{t_L}]^\top\in\mathbb{R}^{L\times 2d}
\end{equation}where $\oplus$ is concatenation operation. Based on such micro-behavior embeddings, two sessions having the same item sequence but different operation sequences still have different representations which can capture users' fine-grainded intentions.

%To add up, we also attempt to directly construct a micro-behavior directed graph, and then directly sends it to the GGNN to obtain the micro-behavior's sequential representation, but the effect is far worse than the result of modeling in this way.(The results can be seen in the experiment part.) There are two main reasons for this failure: (1) The sparseness of the transfer matrix in GGNN is different. The advantage of GGNN lies in capturing the complex transitions in the sequence, but the micro-behavioral matrix is too sparse, and the advantages of GGNN cannot be used well. (2) If the micro-behavior is used to construct the transfer matrix, only the units where two items are switched will constitute a neighbor, and then the representation of other items can be learned, but the item status in other micro-behaviors does not receive neighbor information. In other words, the item information is not effectively transmitted.

\subsubsection{Generating Session Representations}
To obtain a session representation, we should aggregate the embeddings of all micro-behaviors in this session. Inspired by \cite{wu2019session}, we take into account a session's local preference and global preference. A session's local preference is directly represented by the embedding of the most recent micro-behavior, i.e., $\mathbf{m}_L$. 

For representing a session's global preference, we use soft-attention mechanism \cite{softAtt} to assign proper weight for each micro-behavior's embedding in the session since different micro-behaviors have different levels of priority. Specifically, given a micro-behavior $m_t (1\leq t\leq L)$,  its attention weight is computed as
\begin{equation}
        \alpha_t = \boldsymbol{\beta}^\top \sigma(\mathbf{W}_1 \mathbf{m}_L+ \mathbf{W}_2 \mathbf{m}_t + \mathbf{b}_{\alpha})
\end{equation}      
where $\mathbf{b}_{\alpha},\boldsymbol{\beta} \in \mathbb{R}^{2d}$ and $\mathbf{W}_1,\mathbf{W}_2 \in \mathbb{R}^{2d \times 2d}$.
Then, the global representation of the session is
\begin{equation}
       \mathbf{s}_g=\sum_{t=1}^L  \alpha_i \mathbf{m}_t
\end{equation}      
At last, the session's final representation is
\begin{equation}
      \mathbf{s}= \mathbf{W}_3[\mathbf{m}_L;\mathbf{s}_g]\in\mathbb{R}^d
\end{equation}where $\mathbf{W}_3 \in \mathbb{R}^{d \times 4d}$.

After obtaining the representation of session $s$, we compute the final score $\hat{y}_{sj}$ through an MLP fed with $\mathbf{s}$ and the candidate item's embedding $\mathbf{i}_j$, followed by a Softmax operation. Thus we have
\begin{equation}
    \hat{y}_{sj} = softmax\big(MLP(\mathbf{s} \oplus  \mathbf{i}_j)\big)
\end{equation}

To train MKM-SR, we first collect sufficient training samples denoted as $<s, j, y_{sj}>$ where $y_{sj}=1$ if item $j$ is the next interacted item of the user following session $s$, otherwise $y_{sj}=0$. Then we adopt binary cross-entropy as the loss function of SR task as follows,
\begin{equation}\label{eq:Ls}
    \mathcal{L}_S=-\sum\limits_{s\in\mathcal{S}}\sum\limits_{j\in\mathcal{I}}\big\{y_{sj}\log(\hat{y}_{sj}) + (1-y_{sj})\log(1-\hat{y}_{sj})\big\}
\end{equation}where $\mathcal{S}$ and $\mathcal{I}$ are the session set and item set in training samples.
%where $y_{sj}=1$ if denotes the one-hot encoding vector of the ground truth item. Through back-propagation through time algorithm to train the model, we can encode the session information.

\subsection{Learning Knowledge Embeddings}
Recall the toy example of Fig. \ref{fig:micro_behaviors}, song $i_3$ and $i_4$ are the next interacted item of session $s_1$ and $s_2$ respectively. In fact, they are both semantically correlated to the previous items $i_1$ and $i_2$ in terms of shared knowledge (the same singer or genre). As a result, the item embeddings learned based on such shared knowledge are often in consonance with interaction sequences, which are regarded as \emph{knowledge embeddings} in this paper. Such observations inspire us to use knowledge embeddings to enhance SR performance. In this subsection, we introduce how to learn knowledge embeddings given observed item knowledge.

In a KG containing items, many-to-one and many-to-many relations are often observed. For example, many songs are sung by a singer, a movie may belong to several genres and a movie genre includes many movies. Among the state-of-the-art KG embedding models, transH \cite{wang2014transh} imports hyperplanes to handle many-to-many/one relations effectively. Therefore, we import the training loss of TransH to learn knowledge embeddings in our model. 

Specifically, for each attribute relation $r$, we first position a relation-specific translation vector $\mathbf{d}_{r}$ in the relation-specific hyperplane $\mathbf{w}_{r}$. Given a triplet $<i,r,a>$, item $i$'s embedding $\mathbf{i}$ and attribute 
$a$'s embedding $\mathbf{a}$ are first projected to the hyperplane with $\mathbf{w}_{r}$ as the normal vectors. The projections are denoted as $\mathbf{i}_{\perp}$ and $\mathbf{a}_{\perp}$. We expect that $\mathbf{i}_{\perp}$ and $\mathbf{a}_{\perp}$ can be connected by a translation vector $\mathbf{d}_{r}$ on the hyperplane with a low error if $<i,r,a>$ is correct. Thus the score function $\Vert \mathbf{i}_{\perp} + \mathbf{d}_{r} - \mathbf{a}_{\perp} \Vert_2^2$ is used to measure the plausibility that the triplet is incorrect. We can use $\mathbf{w}_{r}$ and $\mathbf{i}$ to represent $\mathbf{i}_{\perp}$ as follows since $\Vert \mathbf{w}_{r} \Vert _2 = 1$.
\begin{equation}
    \mathbf{i}_{\perp} = \mathbf{i} -\mathbf{w}_{r}^\top\mathbf{i}\mathbf{w}_{r}
    \label{eq:h_re}
\end{equation}
Therefore, the loss function for knowledge embedding learning is
\begin{equation}\label{eq:Lk}
    \mathcal{L}_K = \sum_{<i,r,a >\in \mathcal{K}}\Vert (\mathbf{i}-\mathbf{w}_r^\top\mathbf{i}\mathbf{w}_r)+\mathbf{d}_r-(\mathbf{a}-\mathbf{w}_r^\top\mathbf{a}\mathbf{w}_r)\Vert _2^2 
\end{equation}where $\mathcal{K}$ is the set of all knowledge triplets.

\subsection{The Objective of Multi-task Learning}
Many previous recommendation models based on knowledge \cite{DKN,KErec,GANrec} generally learn knowledge embeddings in advance which are used as pre-trained item embeddings. In other words, $\mathcal{L}_K$ is used to pre-train item embedding $\mathbf{i}$ in advance of using $\mathcal{L}_S$ to fine-tune $\mathbf{i}$. In such scenario, knowledge embedding learning and recommendation are two separate learning tasks.

In general, incorporating two learning tasks into an MTL paradigm is more effective than achieving their respective goals separately, if the two tasks are related to each other. In MTL, the learning results of one task can be used as the hints to guide another task to learn better \cite{ruder2017overview}. Inspired by the observations on the example in Fig. \ref{fig:micro_behaviors}, learning knowledge embeddings can be regarded as an auxiliary task to predict the features (item embeddings) which are used for SR's prediction task. Consequently, in MKM-SR we import knowledge embedding learning as an auxiliary task into an MTL paradigm, to assist SR task.

In our scenario, the MTL's objective is to maximize the following posterior probability of our model's parameters $\Phi$ given knowledge triplet set $\mathcal{K}$ and SR's training set $\mathcal{Y}$. %The $\Phi$ includes the embeddings of all items, attributes, attributes relations and other parameters in the neural network . 
According to Bayes rule, this objective is
\begin{equation}
\begin{split}
    &\max p(\Phi | \mathcal{K},\mathcal{Y})= \max \frac{p(\Phi,\mathcal{K}, \mathcal{Y})}{p(\mathcal{K}, \mathcal{Y})} \\ 
    = &\max p(\Phi)  p(\mathcal{K}|\Phi ) p(\mathcal{Y}|\Phi,\mathcal{K})
\end{split}
\label{eq:bayes}
\end{equation}
where $p(\Phi)$ is $\Phi$'s prior probability which is set to follow a Gaussian distribution of zero mean and 0.1 standard deviation. $p(\mathcal{K}|\Phi)$ is the likelihood of observing $\mathcal{K}$ given $\Phi$, and $p(\mathcal{Y}|\Phi,\mathcal{K})$ is the likelihood of observing $\mathcal{Y}$ given $\mathcal{K}$ and $\Phi$, which is defined as the product of Bernoulli distributions. Then, the comprehensive loss function of our MTL's objective is
\begin{equation}\label{eq:total_loss}
	 \mathcal{L} = \mathcal{L}_S + \lambda_1\mathcal{L}_{K} +\lambda_2 \Vert\Phi\Vert_2^2\\ 
\end{equation}where $\Vert\Phi\Vert_2^2$ is the regularization term to prevent over-fitting, and $\lambda_1, \lambda_2$ are control parameters. We obtain the values of $\lambda_1$ and $\lambda_2$ through tuning experiments. 

During the optimization of loss $\mathcal{L}$, there are two training strategies of {MTL, alternating training and joint training \cite{ren2015faster}. For alternating training, we have
\begin{equation}
% \vspace{-0.2cm}
    \centering
	\begin{split}
	& \mathcal{L}_{alter} = -\sum_{s\in\mathcal{S}}\sum_{j\in\mathcal{I}}\big[y_{sj}\log(\hat{y}_{sj}) + (1-y_{sj})\log(1-\hat{y}_{sj})\big] +\\
	 & \lambda_1\sum_{<i,r,a >\in \mathcal{K}} \Vert (\mathbf{i}-\mathbf{w}_r^\top\mathbf{iw}_r)+\mathbf{d}_r-(\mathbf{a}-\mathbf{w}_r^\top\mathbf{aw}_r)\Vert _2^2  +\lambda_2 \Vert\Phi\Vert_2^2\\
	\end{split}
	\label{eq:alter_loss}
\end{equation}where $\mathcal{S}$ and $\mathcal{I}$ represent the set of sessions and candidate items in the training set $\mathcal{Y}$, respectively. %The proposed unit can automatically learn high-order interactions of items in recommender systems and entities in the knowledge.
For joint learning, we have
\begin{equation}
%\vspace{-0.2cm}
    \centering
	\begin{split}
	&\mathcal{L}_{joint} =\sum_{s\in\mathcal{S}}\bigg\{ -\sum_{j\in\mathcal{I}}\big[y_{sj}\log(\hat{y}_{sj}) + (1-y_{sj})\log(1-\hat{y}_{sj})\big] +\\
	 & \lambda_1\sum_{<i,r,a >\in \mathcal{K} \land i\in s} \Vert (\mathbf{i}-\mathbf{w}_r^\top\mathbf{iw}_r)+\mathbf{d}_r-(\mathbf{a}-\mathbf{w}_r^\top\mathbf{aw}_r)\Vert _2^2 \bigg\}  +\lambda_2 \Vert\Phi\Vert_2^2\\
	\end{split}
	\label{eq:joint_loss}
\end{equation}

Through empirical comparisons, we have verified that alternating training is a better strategy for MKM-SR.

\nop{Experiments results show that the multi-loss is effective. In order to understand the reason better, we need to look at the mechanisms that underlie it. Multi-task learning can be regarded as a form of inductive transfer, which help improve a model by introducing an inductive bias, which causes a model to prefer some hypothesis over others. First, as different tasks have different noise patterns, learning the multi task jointly enables the model to obtain a better representation through averaging the noise patterns. Second, learning the knowledge as the auxiliary task provide additional evidence for the relevance and irrelevance of those features. It acts as the regularizer to generalize the model. What' more, through MTL, we can allow the model to eavesdrop. Since the session information can generate good representation for high frequent items and weak for the cold-start items, while knowledge information can learn the cold-start item embedding easily. Besides, the session information can supervise the hyperplane for knowledge to make the knowledge work better for the task.}

\section{Experiments}
%In this section, we first present the experimental datasets, baselines, evaluation metrics and parameter settings. Then we assess the proposed method with compared methods, and compare the performance of different ablation method. Besides, we analyze the performance results with different framework to learn the micro-behavior and the different training phase. 
In this section, we try to answer the following research questions (RQs for short) through extensive experiments.

    \textbf{RQ1:} Can our model MKM-SR outperform the state-of-the-art SR models?
    
    \textbf{RQ2:} Is it useful to incorporate micro-behaviors and knowledge into our model?
   
   \textbf{RQ3:} Is it rational to obtain a session's representation through learning item embeddings by GGNN and learning operation embeddings by GRU separately?
    
    \textbf{RQ4:} Which training strategy is better for incorporating knowledge learning into MKM-SR?
    %\item \textbf{RQ5:} What are the influence of different hyper-parameter settings (the weight of knowledge loss) for our model MAT-SR?
\subsection{Experiment Settings}
\subsubsection{Datasets}
We evaluate all compared models on the following realistic datasets:

\textbf{KKBOX\footnote{https://www.kaggle.com/c/kkbox-music-recommendation-challenge/data}}:  This dataset was provided by a famous music service KKBOX, which contains many users' historical records of listening to music in a given period. We take the 'source system tab' as user operations, such as 'tab my library' (manipulation on local storage) and 'tab search'. The music attributes used in our experiments include artist (singer), genre, language and release year. 

\textbf{JDATA\footnote{https://jdata.jd.com/html/detail.html?id=8}}: This dataset was extracted from JD.com which is a famous Chinese e-commerce website. It contains a stream of user actions on JD.com within two months. The operation types include clicking, ordering, commenting, adding to cart and favorite. The product attributes used in our experiments include brand, shop, category and launch year. 

%\item \textbf{Demo} A fraction 1/100 of JDATA dataset is split based on time for item cold start analysis. And we will slice different rate in later experiments.

%\subsection{Sample Collection}
For both of the two datasets, we considered four item attributes (relations) as knowledge which were incorporated in our model. As in \cite{GRUrec,wu2019session}, we set the duration time threshold of sessions in JDATA to one hour, and set the index gap of sessions in KKBOX to 2000 (according to statistic analysis), to divide different sessions. We also filtered out the sessions of length 1 and the items that appear less than 3 times in the datasets. For both datasets, we took the earlier 90\% user behaviors as the training set, and took the subsequent (recent) 10\% user behaviors as the test set. In model prediction, given a test session the models first computed the matching scores of all items and then generated a top-$k$ list according to the scores.

In order to demonstrate the effectiveness of incorporating item knowledge to alleviate the problem of cold-start items, we added two additional manipulations on our datasets unlike the previous SR evaluations. The first is to retain the items that only appear in the test set, i.e., the cold-start items. The second is to simulate a sparse JDATA dataset, denoted as \textbf{Demo}, through only retaining the early 1\% user behaviors. Such sparse dataset has a bigger proportion of cold-start items. In the previous SR models such as \cite{Liu2018STAMP,wu2019session,MBrec}, these cold-start items' embeddings are initialized in random, and can not be tuned during model training since they are not involved in any training sample. Thus, the recommendations about these items are often unsatisfactory. 

The statistics of our datasets are shown in Table .\ref{tab:datasets}, where '(N)' indicates the datasets having some cold-start items, and 'new\%' is the proportion of the behaviors involving cold-start items to all behaviors in the test set. %The item frequency in table is defined as average item appear times, and equals to the record number divide the item number, which is used to evaluate the sparsity of the dataset. 
We have taken into account all operation types provided by the two datasets. To reproduce our experiment results conveniently, our experiment samples and MKM-SR's source codes have been published on \url{https://github.com/ciecus/MKM-SR}.

%Same as [STAMP] [SRGNN], we use a sequence splitting preprocess that for an input session $s=\{m_1,m_2,\dots,m_n\}$, we generate the sequences and corresponding labels ($[m_1],i_2$),($[m_1,m_2],i_3$), $\dots$,$([m_1,m_2,\dots,m_{n_1}],i_n)$ for training and testing on both datasets, which each input is historical micro-behaviors and label the next interaction item for the current session.
\begin{table}[!htb]
\vspace{-0.1cm}
    \caption{Dataset statistics}    \label{tab:datasets}
    \centering
    \vspace{-0.4cm}
    \begin{tabular}{p{35pt} p{28pt} p{24pt} p{47pt} p{29pt} p{30pt}}
        \toprule
          & seesion\# & session length& item\#(new\%) & item frequency & operation\#\\
        %& KKBOX & KKBOX(N)& JDATA & JDATA(N)\\
        \midrule
        % session\#& 180,047 & 180,096 &455,481 & 456,005\\
        KKBOX & 180,047 & 4.713 & 33,454 & 25.365 & 23 \\
        KKBOX(N) & 180,096 & 4.726 & 34,120(0.99\%) & 24.942 & 23\\
        JDATA & 455,481 & 5.372 & 134,548 & 18.186 & 5\\
        JDATA(N) & 456,005 & 5.383 & 139,099(3.69\%) & 17.654 & 5\\
        Demo & 5,633 & 5.330 & 12,195 & 2.301 & 5 \\
        Demo(N) & 5,696 & 4.992 & 12,917(40.45\%) & 2.192 & 5\\
         %session length&4.713&4.726&5.372&5.383\\
        % item\# (new\%)&33,454&34,124(0.99\%)&134,548&139,099(3.69\%)\\ 
         %item frequency&25.365&24.942&18.186&17.645\\
         %operation\#&23&23&5&5\\
         %\#item attribute&4&4&4&4\\
         \bottomrule
    \end{tabular}
    % \vspace{-0.2cm}
\end{table}

\begin{table*}
    \caption{All models' SR performance scores (percentage value) show that MKM-SR outperforms all competitors no matter whether the historical interactions are sparse.}
    \label{tab:ex_results}
    \centering 
    \vspace{-0.2cm}
    \begin{tabular}
    {p{60pt}p{23pt}p{30pt}|p{23pt}p{30pt}|p{23pt}p{30pt}|p{23pt}p{30pt}|p{23pt}p{30pt}|p{23pt}p{29pt}}
        \toprule
         &\multicolumn{2}{c}{KKBOX}&\multicolumn{2}{c}{KKBOX(N)} &\multicolumn{2}{c}{JDATA}&\multicolumn{2}{c}{JDATA(N)} &\multicolumn{2}{c}{Demo}&\multicolumn{2}{c}{Demo(N)} \\
         &Hit@20 & MRR@20 &Hit@20 & MRR@20 & Hit@20 & MRR@20 & Hit@20 & MRR@20 & Hit@20 & MRR@20 & Hit@20 & MRR@20\\
         \midrule
        FPMC & 5.614 & 1.166 & 5.530&1.147 & 7.531 & 2.623 & 7.049 & 2.493 &3.787 &1.808 &3.261 &1.644 \\
        GRU4REC+BPR & 12.795 & 4.545 & 12.501 & 4.693 & 35.433 & 13.262 & 34.827 & 13.346 &12.189 &5.124 &5.567 & 2.969\\
        GRU4REC+CE & 12.445 & 4.007 &12.429 &4.135 & 35.347 & 13.956 & 34.794 & 13.542 &12.965 &4.992 & 9.896&4.505 \\
        NARM &14.667 & 5.839 & 13.926  &  5.200 & 36.867 & 16.826  & 35.862  &16.677  &  14.446  & 5.645   &  8.056 &3.615      \\
        STAMP & 14.475  & 4.783 & 14.287 & 4.544  &35.555 & 12.936  & 34.691  & 12.187  & 14.609   & 5.796    &9.317 & 2.902   \\
        SR-GNN & 14.187 & 4.476 & 13.399 & 4.792 & 40.588 & 15.968 & 38.723 & 15.203 & 15.504 &7.220 & 10.317 &  4.682\\
        RIB  & 15.982 & 4.763 & 13.887 &5.328 & 37.236  &14.134   & 35.551  & 13.420  & 12.893 &4.887 & 9.965 & 4.436 \\
        \midrule
        KM-SR  & 17.680 & 7.195 & 17.019 & 6.301 & 41.094  & 16.552  & 40.480  & 15.709  & 23.726 &9.363 & 15.065 & 6.323\\       
        M(GRU)-SR &16.971 & 5.435  & 16.865  & 5.250 &37.015 &14.034   & 36.374  & 13.734	  & 18.507 & 6.430	  & 11.262 & 4.747     \\
        {M(GGNN)-SR} & 13.262 & 4.347 & 13.035  & 4.098 & 38.270 & 16.532  & 37.231  & 15.663  & 16.141  & 6.811   &9.168  & 3.572    \\
        {M(GGNNx2)-SR} & 17.270 & 5.532  &16.983  & 5.435 & 41.017  & 16.544  & 41.308  & 15.780  &   19.782 &	7.865  & 12.017  & 	4.734	  \\
        M-SR  & 20.998 & 5.878 & 20.523 &5.707 & 41.440  &   16.851 &41.019   & 15.850  & 20.631 &7.969 & 12.914 & 5.228 \\
        \textbf{MKM-SR}  & \textbf{22.574} & \textbf{7.543} & \textbf{22.221} & \textbf{6.976} & \textbf{42.565} & \textbf{17.585} & \textbf{41.998} & \textbf{16.990} & \textbf{24.623} & \textbf{9.642} & \textbf{15.110} & \textbf{6.424}  \\      
        \bottomrule
    \end{tabular}
\end{table*}

\subsubsection{Compared Models}
To emphasize MKM-SR's superiority performance, we compared it with the following state-of-the-art SR models:

   \textbf{FPMC} \cite{Rendle2010Factorizing}: It is a sequential prediction method based on personalized Markov chain which is often used as SR baseline.
    
    \textbf{GRU4REC+BPR/CE} \cite{hidasi2018recurrent}: These two baselines are the improved versions of GRU4REC \cite{GRUrec} which is a state-of-the-art SR model. GRU4REC+BPR uses Bayes personalized ranking \cite{BPR} as loss function, and GRU4REC+CE uses cross-entropy as loss function.
        
    \textbf{NARM} \cite{Jing2017Neural}: It is a GRU-based SR model with an attention to consider the long-term dependency of user preferences.
    
    \textbf{STAMP} \cite{Liu2018STAMP}: This SR model considers both current interests and general interests of users. In particular, STAMP uses an additional neural network to model current interests.
    
    \textbf{SR-GNN} \cite{wu2019session}: It also utilizes GGNN to capture the complex transition patterns among the items in a session, but does not incorporate micro-behaviors and knowledge.
    
    % \textbf{SR_GNN+K}: It is a variant of SR-GNN in which we concatenate the item embeddings learned by GGNN and the item embeddings learned by TransH.
     
    \textbf{RIB} \cite{MBrec}: It also incorporates user operations of which the embeddings are learned by Word2Vec \cite{W2V}, and adopts GRU to model the sequence of user micro-behaviors. 

In addition, to justify the necessity and validity of incorporating micro-behaviors and knowledge in our model, we further propose some variants of MKM-SR to be compared as follows.

   \textbf{KM-SR}: It removes all modules related to operations and the rest components are the same as MKM-SR. We compared MKM-SR with KM-SR to verify the significance of incorporating micro-behaviors. 
   
    \textbf{M-SR}: It removes the auxiliary task of learning knowledge embeddings, i.e., $\mathcal{L}_{K}$ in Eq. \ref{eq:total_loss}, and the rest components are the same as MKM-SR. All of the following variants remove the task of learning knowledge embeddings, between which the differences are the manipulations on session modeling.
        
    \textbf{M(GRU/GGNN)-SR}: Unlike MKM-SR, these two variants directly learn micro-behavior embeddings ($\mathbf{m}_t$). The only difference between them is, M(GRU)-SR feeds micro-behavior sequences to GRU and M(GGNN)-SR feeds micro-behavior sequences to GGNN.
    
     \textbf{M(GGNNx2)-SR} It uses two GGNNs to learn operation embeddings and item embeddings respectively.

\subsubsection{Evaluation Metrics}
We use the following metrics to evaluate all models' performance which have been widely used in previous SR evaluations.

    \textbf{Hit@k}: It is the proportion of hit samples to all samples that have the correct next interacted item in the top-$k$ ranking lists.
    
    \textbf{MRR@k}: The average reciprocal rank of the correct next interacted item in the top-$k$ ranking list. The reciprocal rank is set to zero if the correct item is ranked behind top-$k$.
    
\subsubsection{Hyper-parameter Setup}
For fair comparisons, we adopted the same dimension of operation and item embeddings for MKM-SR and all baselines. Due to space limitation, we only display the results of 100-dim embeddings. The consistent conclusions were drawn based on the experiment results of the embeddings of other dimensions. In addition, all embeddings were initialized by a Gaussian distribution with a mean of 0 and a standard deviation of 0.1. We set GGNN's step number $H$ to 1. MKM-SR was learned by alternating training rather than joint training, of which the reason will be verified in the following experiments. In addition, we used Adam \cite{Adam} optimizer with learning rate 0.001 and batch size 128. %In particular, we use batch normalization between dot-attention layer and session-attention layer to prevent overfitting. 
For the baselines, we used their default hyper-parameter settings in their papers except for embedding dimension. About the control parameters in Eq. \ref{eq:total_loss}, we set $\lambda_1$ in Eq. \ref{eq:total_loss} to 0.0001 for each dataset, which was decided through our tuning experiments. For L2 penalty $\lambda_2$, we set it to $10^{-5}$ as previous SR models \cite{Wang2019SurveySession}.

\vspace{0.2cm}
Next, we display the results of our evaluations to answer the aforementioned RQs, based on which we provide some insights on the reasons causing the superiority or inferiority of compared models .

\subsection{Global Performance Comparisons}
At first, we compared all models' SR performance over different datasets to answer RQ1, of which the Hit@20 and MRR@20 scores (percentage value) are listed in Table \ref{tab:ex_results}. The displayed scores are the average of five runnings for each model. 

The comparison results show that, MKM-SR outperforms all baselines and variants in all dataset (answer yes to RQ1). Especially in the datasets with cold-start items and Demo, MKM-SR and KM-SR have more remarkable superiority. Such results justify the effects of incorporating knowledge to alleviate the sparsity problem of cold-start items (answer yes to RQ2). As shown in Table \ref{tab:datasets}, KKBOX has more unique operations than JDATA which are useful to better capture user preferences on fine-grained level. Therefore, besides MKM-SR and M-SR, another model incorporating user operations RIB also has more remarkable advantage in KKBOX than in JDATA, compared with the GRU-based baselines that do not incorporate operations. These results justify the effects of incorporating micro-behaviors (answer yes to RQ2). 

In addition, a user is more likely to interact with the same item in a session of JDATA. The transition pattern between the successive items in such scenario can be captured by GGNN better than GRU. It is the reason that SR-GNN has greater advantage in JDATA than in KKBOX, compared with the GRU-based models including GRU4REC+BPR/CE and NARM. 

\subsection{Ablation Study}
We further compared MKM-SR with its variants to answer RQ2, RQ3. We have the following conclusions drawn based on the results in Table \ref{tab:ex_results}. MKM-SR's advantage over KM-SR and M-SR shows that, both micro-behaviors (operations) and item knowledge deserve to be incorporated w.r.t. improve SR performance (answer yes to RQ2). In addition, M-SR outperforms M(GRU)-SR and M(GGNN)-SR indicating that modeling a session through learning item embeddings and learning operation embeddings separately is more effective than learning micro-behavior embeddings directly. As we stated before, the transition pattern of item sequences is different to that of operation sequence. Therefore, it is less effective to combine an item with an operation as a micro-behavior and then learn the micro-behavior sequence only by a certain model. Furthermore, M-SR's superiority over M(GGNNx2)-SR shows that operation sequences should be learned by GRU rather than GGNN, of which the reason has been explained in Subsec. \ref{sec:session}. These results provide yes answer to RQ3.

\subsection{Strategies of Incorporating Knowledge Learning}
As we mentioned before, there are two training strategies for our MTL loss Eq. \ref{eq:total_loss}, i.e., alternating training (Eq. \ref{eq:alter_loss}) and joint training (Eq. \ref{eq:joint_loss}). To answer RQ4, we trained KM-SR respectively with the two training strategies and compared their learning curves. Furthermore, we added a pre-training variant to be compared, in which the item embeddings are first pre-trained by TransH and then input into KM-SR to be tuned only by loss $\mathcal{L}_S$ in Eq. \ref{eq:Ls}. We did not adopt MKM-SR in this comparison experiment because the three training strategies are not relevant to operation embedding learning.

\begin{figure}[htbp]
    \centering
    \includegraphics[width=8.5cm]{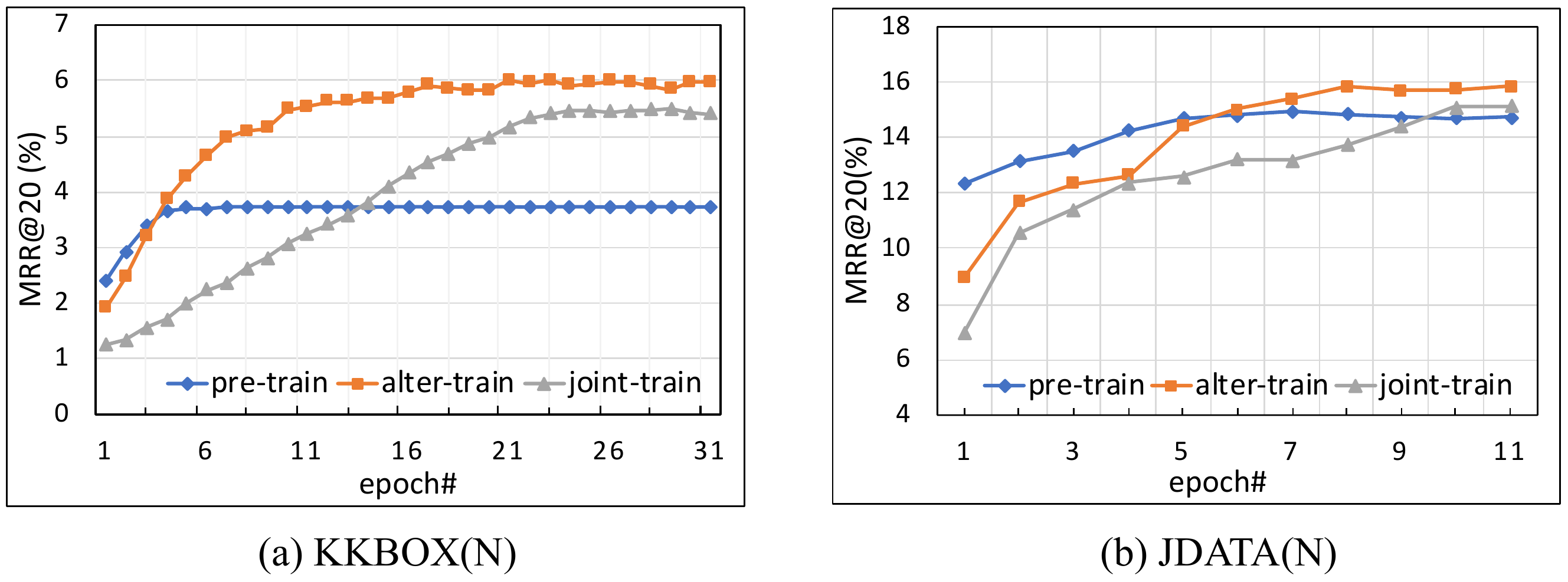}
    \vspace{-0.5cm}
    \caption{The learning curves of three different strategies to incorporate knowledge learning show that, incorporating knowledge learning into the MTL of alternating training is the best strategy for our SR task.} \label{fig:train_cpr}
       %\vspace{-0.2cm}
\end{figure}

In Fig. \ref{fig:train_cpr}, we only display the learning curves of MRR@20 in KKBOX(N) and JDATA(N) since MRR@k reflects ranking performance better than Hit@k. The curves in the figure show that, although the pre-training model has a better learning start, it is overtaken by the other two rivals on the stage of convergence. Such results demonstrate MTL's superiority over the knowledge-based pre-training. According to Eq. \ref{eq:joint_loss}, the items often occurring in the sessions of training set will be tuned multiple times by loss $\mathcal{L}_K$ in each epoch of joint training. It makes the learned embeddings bias to the auxiliary task $\mathcal{L}_K$ too much, shrinking the learning effect of the main task $\mathcal{L}_S$. Therefore, alternating training is better than joint training in our SR scenario.
\begin{figure}[htbp]
\vspace{-0.2cm}
    \centering
    \includegraphics[width=9cm]{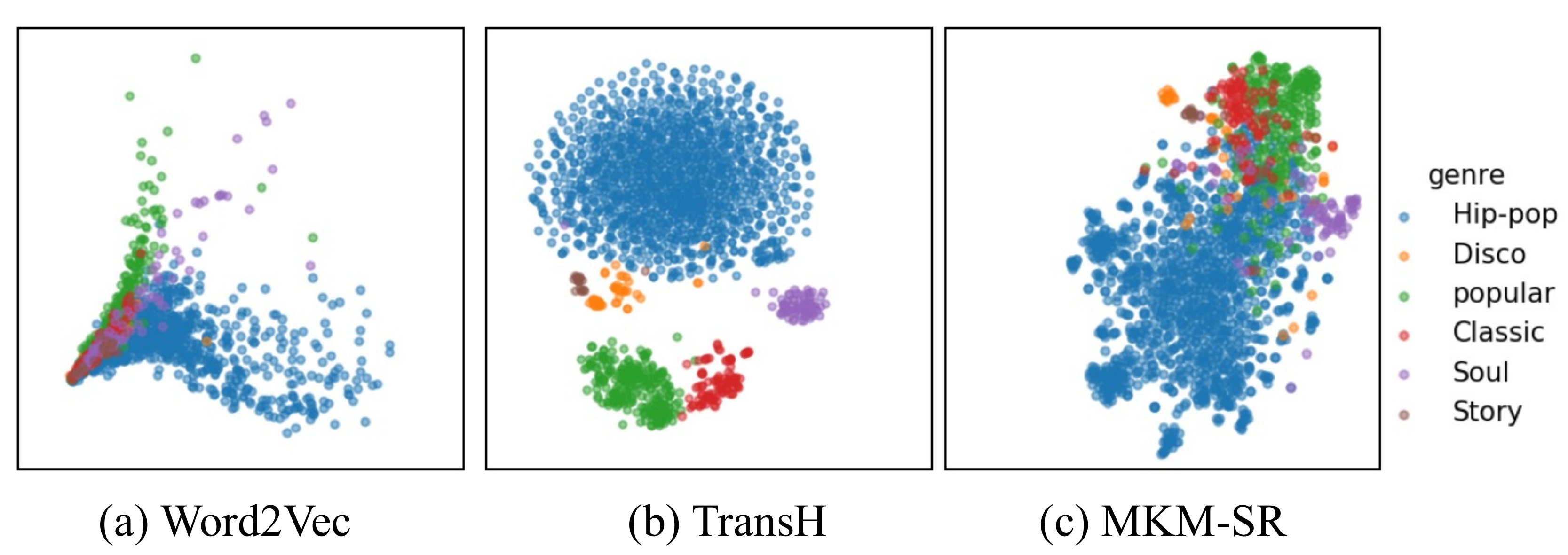}
    \vspace{-0.5cm}
    \caption{KKBOX(N)'s item embedding distributions under different learning mechanisms show that, the song embeddings learned by the MTL keep close distances across different groups (genres) on the premise of discriminating different groups. It conforms to the fact that the successive songs in a session may belong to different genres.} \label{fig:item_ebd}
      % \vspace{-0.2cm}
\end{figure}

To further verify the significance of incorporating knowledge learning into the MTL paradigm (Eq. \ref{eq:total_loss}), we visualize the embedding distributions of some items sampled from KKBOX(N), which were learned respectively by different mechanisms in Fig. \ref{fig:item_ebd} where the points of different colors represent the songs of different genres. In Fig. \ref{fig:item_ebd}(a), the item embeddings were learned by feeding item sequences of sessions into Word2Vec, thus two items are close in the space if they often cooccur in some sessions. As shown in the sub-figure, such learned embeddings make many songs of different genres too converged and thus can not discriminate different genres. In Fig. \ref{fig:item_ebd}(b), the item embeddings were learned solely by TransH. Although such learned embeddings discriminate different genres obviously, the gap between two groups of different genres is too big. It makes the model based on embedding distances hard to predict the item of different genre as the next interacted item, which does not conform to some facts, such as the song $i_3$ in $s_1$'s item sequence shown in Fig. \ref{fig:micro_behaviors}. It is also the reason why the pre-training model is defeated by the joint-training model and alter-training model in Fig. \ref{fig:train_cpr}. The item embeddings shown in Fig. \ref{fig:item_ebd}(c) were learned by MKM-SR through MTL, and exhibit two characteristics, i.e., they can discriminate different genres for most items, meanwhile keep close distances across different genres. Such item embeddings with the two characteristics well indicate two kinds of correlations between the successive items in a session. The former characteristic indicates the semantic correlations among items, and the latter characteristic indicate the items' co-occurrence correlations across different sessions. In fact, these two correlations can be captured respectively through the learning task of $\mathcal{L}_K$ and the learning task of $\mathcal{L}_S$. Obviously, it is useful for improving SR to capture these two correlations simultaneously.
 
\subsection{MTL's Control Parameter Tuning}
At last, we investigate the influence of $\lambda_1$'s in the MLT's loss $\mathcal{L}$ to MKM-SR's final recommendation performance. Fig. \ref{fig:lambda} shows MKM-SR's MRR@20 scores in KKBOX(N), from which we find that MKM-SR's performance varies marginally ($\sim$1\%) when $\lambda_1$ is set in $[0.00001,0.1]$. What's more, MKM-SR gains the best score when $\lambda_1=0.0001$. It implies that, as an auxiliary task knowledge embedding learning will disturb the main task of SR if it is assigned with more weight. %second, for sparse and cold start item, the weight is enough for them to be adapt through iteration.
 \begin{figure}[htbp]
\vspace{-0.1cm}
    \centering
    \includegraphics[width=4.6cm]{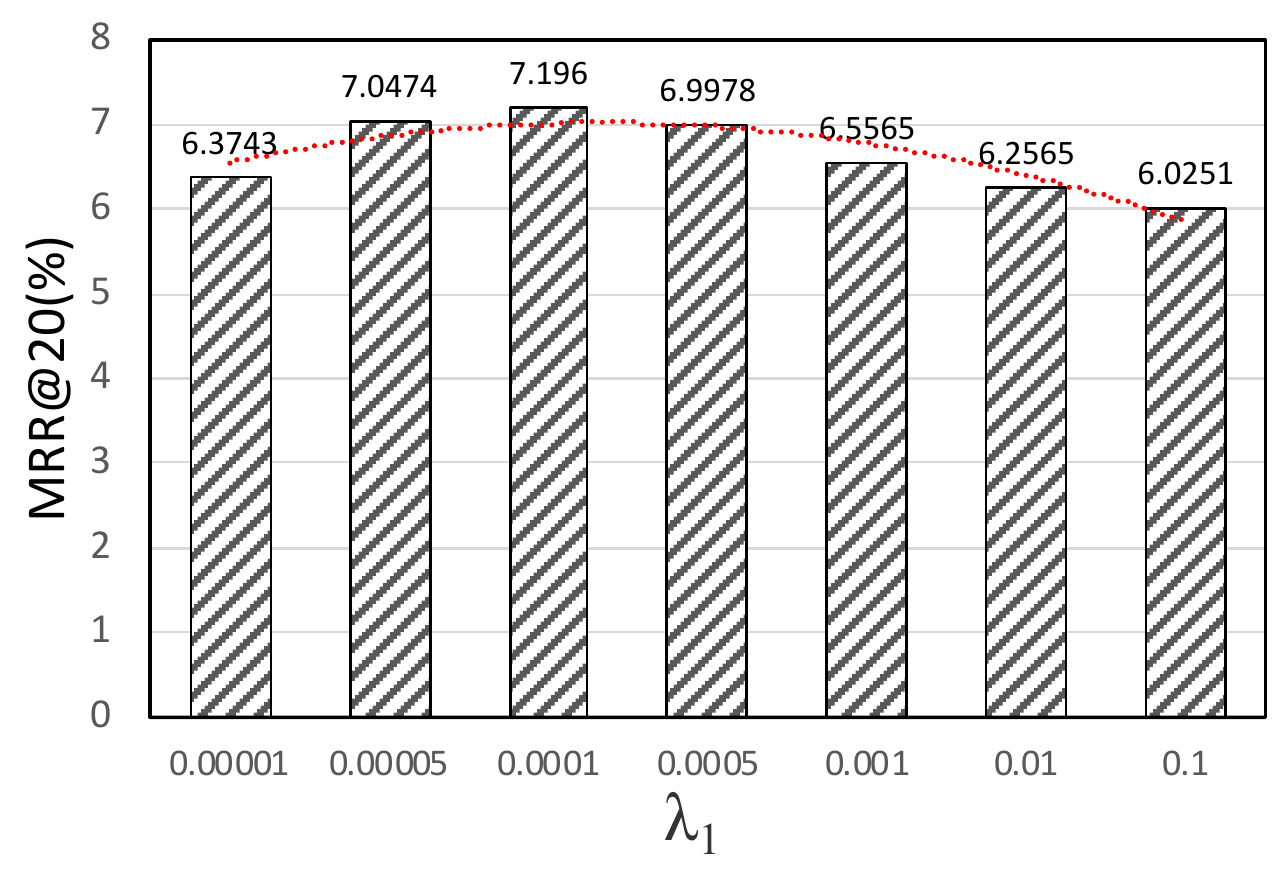}
    \vspace{-0.4cm}
    \caption{MKM-SR's performance on KKBOX(N) with different $\lambda_1$ shows that $\lambda_1=0.0001$ is the best setting.} \label{fig:lambda}
       \vspace{-0.2cm}
\end{figure}

%\vspace{-0.2cm}
\section{Conclusion}
In this paper, we propose a novel session-based recommendation (SR) model, namely MKM-SR, which incorporates user micro-behaviors and item knowledge simultaneously. According to the different intuitions about item sequences and operation sequences in a session, we adopt different mechanisms to learn item embeddings and operation embeddings which are used to generate fine-grained session representations. We also investigate the significance of learning knowledge embeddings and the influences of different training strategies through sufficient comparison studies. MKM-SR's superiority over the state-of-the-art SR models is justified by our extensive experiments and inspires a promising direction of improving SR.

\bibliographystyle{ACM-Reference-Format}
\bibliography{reference}
\end{document}